\documentclass[aps,prb,twocolumn,showpacs,superscriptaddress]{revtex4}
\usepackage{graphicx}
\usepackage{color}

\renewcommand{\vec}[1]{\mathbf{#1}}
\begin{document}

\preprint{APS/123-QED}

\title{Three Dimensional Edwards-Anderson Spin Glass Model in an External Field}% Force line breaks with \\

\author{Sheng~Feng}
\affiliation{%
 Department of Physics and Astronomy, Louisiana State University, Baton Rouge, LA 70803, USA
}
 \affiliation{%
Center for Computation and Technology, Louisiana State University, Baton Rouge, LA 70803, USA
}
\author{Ye~Fang}%
\affiliation{%
Center for Computation and Technology, Louisiana State University, Baton Rouge, LA 70803, USA
}
\affiliation{%
ECE Division, School of Electrical Engineering and Computer Science, Louisiana State University, Baton Rouge, LA 70803, USA
}
\author{Ka-Ming~Tam}
\affiliation{%
 Department of Physics and Astronomy, Louisiana State University, Baton Rouge, LA 70803, USA
}
 \affiliation{%
Center for Computation and Technology, Louisiana State University, Baton Rouge, LA 70803, USA
}

\author{Zhifeng~Yun}
\affiliation{%
Center for Computation and Technology, Louisiana State University, Baton Rouge, LA 70803, USA
}
\affiliation{%
Center for Advanced Computing and Data Systems, University of Houston, Houston, TX 77204, USA}

\author{J.~Ramanujam}
\affiliation{%
Center for Computation and Technology, Louisiana State University, Baton Rouge, LA 70803, USA
}
\affiliation{%
ECE Division, School of Electrical Engineering and Computer Science, Louisiana State University, Baton Rouge, LA 70803, USA
}

\author{Juana~Moreno}
\affiliation{%
 Department of Physics and Astronomy, Louisiana State University, Baton Rouge, LA 70803, USA
}
 \affiliation{%
Center for Computation and Technology, Louisiana State University, Baton Rouge, LA 70803, USA
}

\author{Mark~Jarrell}
\affiliation{%
 Department of Physics and Astronomy, Louisiana State University, Baton Rouge, LA 70803, USA
}
 \affiliation{%
Center for Computation and Technology, Louisiana State University, Baton Rouge, LA 70803, USA
}

\date{\today}% It is always \today, today,

\begin{abstract}

We study the Edwards-Anderson model on a simple cubic lattice with a finite 
constant external field. We employ an indicator composed of a ratio of 
susceptibilities at finite wavenumbers, which was recently proposed to avoid 
the difficulties of a zero momentum quantity, for capturing the spin glass 
phase transition. Unfortunately, this new indicator is fairly noisy, so 
a large pool of samples at low temperature and small external field are 
needed to generate results with sufficiently small statistical error for 
analysis.  We thus implement the Monte Carlo method using graphics 
processing units to drastically speedup the simulation. We confirm previous 
findings that conventional indicators for the spin glass transition, including 
the Binder ratio and the correlation length do not show any indication of a 
transition for rather low temperatures.  However, the ratio of spin glass 
susceptibilities do show crossing behavior, albeit a systematic analysis 
is beyond the reach of the present data. This calls for a more thorough study 
of the three dimensional Edwards-Anderson model in an external field.

\iffalse
\begin{description}
\item[Usage]
Secondary publications and information retrieval purposes.
\item[PACS numbers]
May be entered using the \verb+\pacs{#1}+ command.
\item[Structure]
You may use the \texttt{description} environment to structure your abstract;
use the optional argument of the \verb+\item+ command to give the category of each item. 
\end{description}
\fi
\end{abstract}

\maketitle

\textit{Introduction.} Most spin systems order when the temperature is sufficiently low. 
Conventional magnetic orderings break spin symmetry, and the moments align 
in a pattern with long range order. However, magnetic 
systems with random frustrated couplings can avoid conventional 
ordering by breaking ergodicity. Typical spin glass systems with 
such competing magnetic couplings include localized spins in metals coupled 
via the oscillating Rudermann-Kittel-Kasuya-Yoshida 
exchange as CuFe and CuMn, and in insulators with competing interactions 
as in LiHoYF and EuSrS \cite{Binder-Young1986,Mydosh-1993,Diep-2004}. These 
systems do not display long range order for a wide range of diluted spin 
concentrations.

A widely studied model to describe spin glass physics is the Edwards-Anderson 
(EA) model\cite{Edwards-Anderson-1975}. It is composed of spins interacting
with their nearest neighbors via random couplings. The mean-field variant of the 
EA model, the Sherrington-Kirkpatrick (SK) model\cite{Sherrington-Kirkpatrick1978,Sherrington-Kirkpatrick-1975},
was solved by the replica technique in 1975 with the striking observation that 
the entropy can be negative at low temperature\cite{Sherrington-Kirkpatrick-1975,Sherrington-Kirkpatrick1978}. 
A cavity mean field method was proposed by Thouless, Anderson 
and Palmer (TAP) in which the local magnetization of each site is considered as 
an independent order parameter\cite{Thouless-Anderson-Palmer-1977}. The hope was to 
obtain a more physical mean field solution without involving the replica technique. 
However, multiple solutions were found\cite{Bray-Moore-1980}.

Motivated by the deficits of previous approaches, de Almedia and Thouless
further studied the replica symmetric mean field solution and found a line 
in the temperature--magnetic field plane where the replica symmetry
solution is unstable towards replica symmetry breaking (RSB) \cite{Almedia-Thouless-1978}. 
The replica overlap has more structure than simply a constant. The way to 
characterize this structure for a stable mean field solution was developed 
by Parisi \cite{Parisi-1980a,Parisi-1980b,Parisi-1980c}.  There is a hierarchy 
of the replica overlap, and this can be described in terms of a ultra-metric tree. 
The replica symmetry breaking scheme resolved the negative entropy crisis and naturally 
explained the many solutions found in the TAP approach.

The RSB theory is accepted to be the correct description of the SK model, indeed 
it provides the exact free energy \cite{Talagrand-2006,Guerra-2003}. 
However, its applicability to real spin glasses has been intensively debated 
over the last three decades, especially in the three dimensions case. For systems 
below the upper critical dimension \cite{Harris-Lubensky-Chen-1976,Tasaki-1989,Green-Moore-Bray-1983} 
the most prominent competing theory is the droplet model elaborated by 
Huse and Fisher \cite{Fisher-Huse-1988,Fisher-Huse-1987} and based on the idea of 
domain wall scaling by Moore, Bray and McMillan \cite{McMillan-1984,Bray-Moore-1987}. 
In this theory, there exists a finite characteristic length scale where droplets of 
excitations can loose energy by aligning with the field. The spin glass phase is thus 
destroyed by any finite external field. Moreover, those excitations are assumed to 
be compact and with fractal dimension smaller than the spacial dimension, in contrast 
with the space-filling excitations in the mean field theory. 

Thus a possible scheme to discern between the RSB and the droplet theories is to determine
whether a spin glass phase exists at a finite external field \cite{Young-Katzgraber2004}. 
There are other schemes based on the differences in the overlap and the excitations in these two theories. 
For example, the distribution of the overlap and parameters that characterize it \cite{Namoichi-Gubernatis-2002, % what is "it"
Marinari-etal-1998,Marinari-etal-1999,Bokil-etal-1999,Moore-etal-1998,Monthus-Garel-2013}, 
the existence of the ultra-metric structure in the overlap \cite{Hed-Young-Domany-2004,Contucci-etal-2007}, 
and the nature of the ground state and its 
excitations \cite{Palassini-Young-2000a,Palassini-Young-2000b,Aspelmeier-Moore-Young-2003,Marinari-Parisi-2001,
Houdayer-Martin-1999,Marinari-Parisi-Zuliani-2000,Marinari-etal-1999}.
Unfortunately, the conclusions draw from different studies are often controversial.  
This is mostly due to two factors, the limitation in the system sizes that can be 
simulated and the interpretation of the data. 

Using the same techniques on the three dimensional EA model under an external field, no signal of 
a crossing of the scaled correlation length for different system sizes can be 
detected\cite{Young-Katzgraber2004}.  We will show this is also 
the case for the Binder ratio.  The absence of crossing is a powerful evidence that a spin
glass phase is absent in the presence of an external field. However, it has been 
argued that the system sizes studied may be too small and far from the scaling 
regime. To remedy this problem, one dimensional models with long range power-law 
decaying interactions \cite{Kotliar-Anderson-Stein-1983} which mimic the short range models 
at higher dimensions have been intensively studied over last few years \cite{Katzgraber-Young-2003a,Katzgraber-Young-2003b,
Leuzzi-1999}. In these models much larger systems can be studied \cite{Katzgraber-Larson-Young-2009,
Katzgraber-Hartmann-2009,Leuzzi-etal-2008,Larson-etal-2013}.

On top of these controversies, it has been recently argued that the scaled correlation length 
is not a good parameter for the spin glass transition in a field since its calculation involves 
the susceptibility at zero momentum \cite{Leuzzi-etal-2008}.
%Therefore a better indicator should avoid the zero momentum susceptibility. 
The latest proposal is to study the ratio of susceptibilities at the 
two smallest non-zero momenta, denoted it as $R_{12}$ \cite{Banos-2012}. It has 
been shown that in four dimensions this quantity displays a crossing
at finite temperature which is an important clue that the spin glass can still 
exist without time reversal symmetry below the upper critical dimension \cite{Banos-2012}. 
Giving the success of using $R_{12}$ to capture the spin glass phase
at four dimensions, we reexamine the three dimensional EA model on a simple cubic lattice using 
a new development in computer architecture, and the recently proposed $R_{12}$. 
We will demonstrate that graphic card computing is particularly well suited for 
equilibrium simulations of spin glass systems, in particular for cases where a huge number 
of realizations is required such as the model we study in this work. 

\textit{Methods and Measured Quantities.} The Hamiltonian for the EA model is given as
\begin{eqnarray}
H=-\sum_{<i,j>} J_{ij} S_{i}S_{j}-h\sum_{i}S_{i},
\label{Hamiltonian}
\end{eqnarray}
where $S_i$ indicate Ising spins on a simple cubic lattice with $N=L^3$ sites and periodic boundary conditions. 
The coupling $J_{ij}$ is bimodal distributed with probability %function 
$P(J_{ij}) = \frac{1}{2}(\delta(J_{ij}-1) + \delta(J_{ij}+1))$, and $h$ is an external field.

%\subsection{Observables}
The spin glass overlap is defined as 
\begin{equation}
  \label{eq:overlap}
  q(\vec{k})=\frac{1}{N}\sum_{j}S_j^{(\alpha)} S_j^{(\beta)}\exp^{i\vec{k} \cdot \vec{r}_j},
\end{equation}
where $\alpha$ and $\beta$ are two independent realizations of the same disorder model.
We calculate the overlap kurtosis or the Binder ratio from the overlap as \cite{Ciria-etal-1993,Marinari-etal-1998}
\begin{equation}
  \label{eq:binder}
  g=\frac{1}{2}\left(3-\frac{\overline{\left<\left(q(0)-\overline{\left<q(0)\right>}\right)^4\right>}}{\overline{\left<\left(q(0)-\overline{\left<q(0)\right>}\right)^2\right>}^2}\right).
\end{equation}
Note that $\overline{(\cdots)}$ indicates averaging over different disorder realizations,
and $\left<\cdots\right>$ denotes thermal averaging.

The wave vector dependent spin glass susceptibility is defined as \cite{Marinari-etal-1998}
\begin{equation}
  \label{eq:chi}
  \chi(\vec{k})= N(\overline{\left<q^2(\vec{k})\right>}-\overline{\left<q(\vec{k})\right>}^2),
\end{equation}
and the correlation length as 
\begin{equation}
  \label{eq:corr}
  \xi_L=\frac{1}{2\sin(\vec{k}_{\mathrm{min}}/2)}\left[\frac{\chi(0)}{\chi(\vec{k}_{\mathrm{min}})}-1\right]^{1/2},
\label{eq:corrlength}
\end{equation}
where $\vec{k}_{\mathrm{min}}=(2\pi/L,0,0)$. 

We define $R_{12}$ as the ratio between the susceptibilities with the two smallest non-zero wave 
vectors \cite{Banos-2012}
\begin{equation}
  \label{eq:r12}
  R_{12}=\frac{\chi(\vec{k}_1)}{\chi(\vec{k}_2)},
\end{equation}
where $\vec{k}_1=(2\pi/L,0,0)$, $\vec{k}_2=(2\pi/L,2\pi/L,0)$.

Parallel tempering\cite{Hukushima-Nemoto1996,Marinari-Parisi1992} is used to accelerated the thermalization, 
in which $N_T$ samples of the same disorder coupling are simulated in parallel within a range 
of temperatures. In order to compute 
the spin glass overlap (Eq.\ref{eq:overlap}) we simulate two replicas 
of the system with the same bonds $J_{ij}=\pm 1$ and field $h$ at each temperature. 

We implement the Monte Carlo simulation with parallel tempering on graphics 
processing units using the CUDA programming language \cite{Nickolls:2008:SPP:1365490.1365500}. 
Multispin coding\cite{PhysRevLett.42.1390,Zorn1981337} is used 
to pack the $N_T$ replicas into the small but extremely fast shared 
memory. We achieve a performance of 33ps per spin flip attempt on a GTX 580 card.
We use the CURAND implemented XORWOW generator to generate random numbers \cite{curand}.
Since the GPU is a commodity hardware and widely available in large
computer clusters, it is now easy to greatly accelerate these 
simulations. The details of the implementation can be found in Ref \onlinecite{Fang-Feng-Tam-etal-2013}. 
%\cite{Zhang:2012:AAS:2259016.2259037,Weigel:2012:PPS:2151219.2151631,preis2011gpu,Preis:2009:GAM:1537305.1537344,Nguyen:2010:BOS:1884643.1884658,Micikevicius:2009:FDC:1513895.1513905,Maruyama:2011:PIP:2063384.2063398,doi:10.1142/S0129183112400025,DBLP:journals/ijpp/HawickLP11,CSTN-093,2012JCoPh.231.1209K,2012arXiv1204.6193M,2012arXiv1204.6192Y,2011CoPhC.182.1833W,2011arXiv1107.5463W,2010CoPhC.181.1549B,2010arXiv1006.2566B}

\begin{table}[ht]
  \centering
  \caption{Parameters of the simulations.  $L$ is the linear system size.
$N_{\mathrm{samp}}$ is the number of samples, 
$N_{\mathrm{sweep}}$ is the total number of Monte Carlo sweeps for each of the 2$N_T$ replicas 
for a single sample, $\beta_{\mathrm{max}}$ and $\beta_{\mathrm{min}}$ shows the temperature 
region simulated, and $N_T$ is the number of temperatures used in the parallel tempering method. 
The temperature set in each simulation follows a geometric distribution, 
i.e. $\beta_n=\beta_{\rm min}\alpha^{n-1}$, where $\alpha=(\beta_{\rm max}/\beta_{\rm min})^{1/(N_T-1)}$, $n\in [1,N_T]$.
The first half of the Monte Carlo sweeps are used for thermalization and the second half are used for measurement.}
 \begin{ruledtabular}
 \begin{tabular}{lrrrrr}
    $L$&$N_{\mathrm{samp}}$&$N_{\mathrm{sweep}}$& $N_T$  &$\beta_{\mathrm{max}}$ &  $\beta_{\mathrm{min}}$  \\
\hline
%    4&190,000&2,000,000&24&2.0&0.7\\
6&500,000&2,000,000&56&1.8&0.1\\
8&350,000&2,000,000&56&1.8&0.1\\
10&240,000&2,000,000&56&1.8&0.1\\
%12&30,000&20,000,000&0.7&1.8&24\\
%16&30,000&20,000,000&0.7&1.5&24\\
  \end{tabular}
\end{ruledtabular}
  \label{tab:parameters}
\end{table}

We list the parameters of our simulation in Table \ref{tab:parameters}. We 
benchmarked the code against existing results at $h=0$.  The smallest 
$\beta$ used in the parallel tempering is well below the critical temperature ($1/\beta_{c} = T_{c} \approx 1.1019 \pm 0.0029$) \cite{Baity-Jesi-etal-2013} of 
the spin glass transition at zero field\cite{Ballesteros2000,Baity-Jesi-etal-2013}, while the largest $\beta$ 
is about two times larger.  The estimated critical field at zero temperature is around $h \approx 0.65$ for 
the model with zero mean and unit variance Gaussian distributed couplings\cite{Krzakala-etal-2001}. 
We choose to work in a relatively small field, $h=0.1$. 
%All the data we show use this strength of field. 
The jackknife method is used to estimate the statistical errors from disorder averaging.

{\it Results.} We plot the spin glass susceptibility in Fig. \ref{fig:Chi}. As in the 
zero field case, the susceptibility increases as the temperature is lowered, however there is
no obvious asymptotic scaling behavior. In particular, for temperatures below  
the zero-field critical temperature, the
slope of the curves decreases and they begin to bend downward. This result is similar to the one obtained for
the one dimensional model\cite{Larson-etal-2013}, but in contrast with the results of the four dimensional lattice 
which displays asymptotic divergent susceptibilities \cite{Marinari-etal-1998}.

\begin{figure}[ht]
  \includegraphics[width=0.5\textwidth]{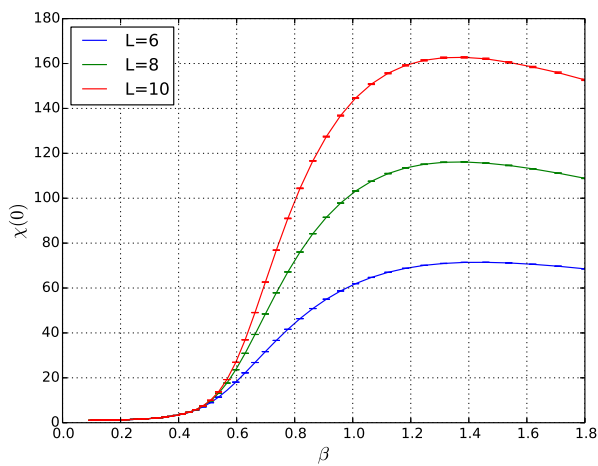}% Here is how to import EPS art
  \caption{\label{fig:Chi} Spin glass susceptibility at zero momentum, $\chi(0)$, as a function of inverse temperature 
for system sizes $L=6,8,10$.}
\label{fig:Chi}
\end{figure}

\begin{figure}[ht]
  \includegraphics[width=0.5\textwidth]{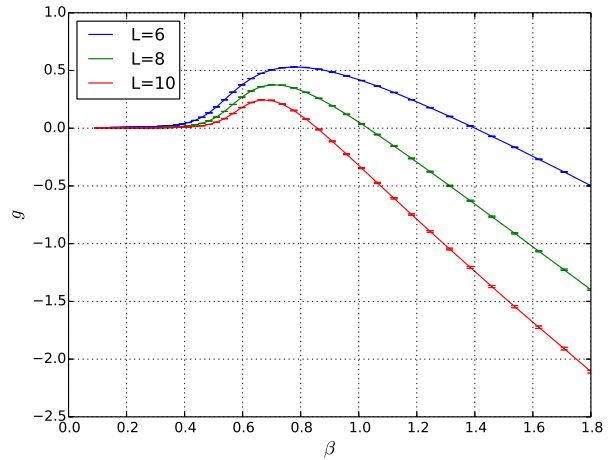}% Here is how to import EPS art
  \caption{\label{fig:b-h0} Binder ratio as a function of inverse temperature in the
range  $\beta=0.1\sim1.8$ for system sizes $L=6,8,10$.}
\label{fig:Binder}
\end{figure}

As the susceptibility does not show a behavior in accordance with the conventional 
finite size scaling theory for a second order transition, we move to study various 
cumulants and ratios of susceptibilities of the overlap parameter. We show the Binder ratio in the Fig.~\ref{fig:Binder}. It does not display any signal of crossing. 
Indeed, the curves for different system sizes do not even tend to merge as the temperature 
is lowered.  Note that the Binder ratio corresponds to the fourth-order cumulant of the 
distribution, and 
%spin glass susceptibility at zero momentum. 
the possible issues related with the soft mode  
contributing to the zero momentum susceptibility should likely be canceled in the Binder ratio.

\begin{figure}[ht]
  \includegraphics[width=0.5\textwidth]{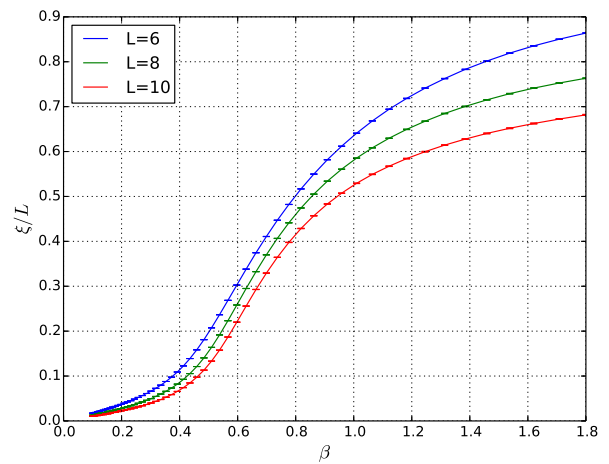}% Here is how to import EPS art
  \caption{\label{fig:c-h0} Scaled correlation length $\xi/L$ as a function of inverse temperature for system sizes $L=6,8,10$.  }
\end{figure}

Fig.~\ref{fig:c-h0} displays the scaled correlation length. This is now a standard 
diagnosis for the detection of a spin glass transition. The correlation length is 
extracted from the Ornstein-Zernike form (Eq.~\ref{eq:corrlength}), and thus essentially given by the ratio 
between the zero and the smallest finite momentum susceptibilities. 
Similar to the Binder ratio, and consistent with other results in the literature, 
there is no crossing or even merging down to a rather low temperature~\cite{Young-Katzgraber2004}.

From now on we focus on $R_{12}$. We first perform simulations in zero field
where $R_{12}$ shows a crossing close to the expected critical temperature found from
the Binder ratio and the correlation length. Therefore, the crossing in $R_{12}$ should 
be a viable indicator for the phase transition. Unfortunately, we find that $R_{12}$ is in 
general much nosier than other quantities. This is due to the fact that the sampling 
of higher momentum quantities is almost always characterized by larger statistical 
fluctuations. Taking the ratio between two susceptibilities at finite momenta clearly further
harms the quality of the data. To reduce the error bars we generate long runs and larger 
pools of disorder realizations (see Table \ref{tab:parameters}). This is the main reason we have generated a rather large number ($2.4 \times 10^5$) of realizations 
for the largest systems size we present here, and even more for smaller sizes. To further reduce the 
fluctuations, we impose all point group symmetries. For example, when we calculate 
$\chi(2\pi/L,0,0)$ we average the susceptibility at three different directions 
($\chi(2\pi/L,0,0)$, $\chi(0,2\pi/L,0)$, and $\chi(0,0,2\pi/L)$). This averaging implicitly assumes 
that the point group symmetry is restored which is justified only when the number of realization
is rather large. 
 
\begin{figure}[ht]
\includegraphics[width=0.5\textwidth]{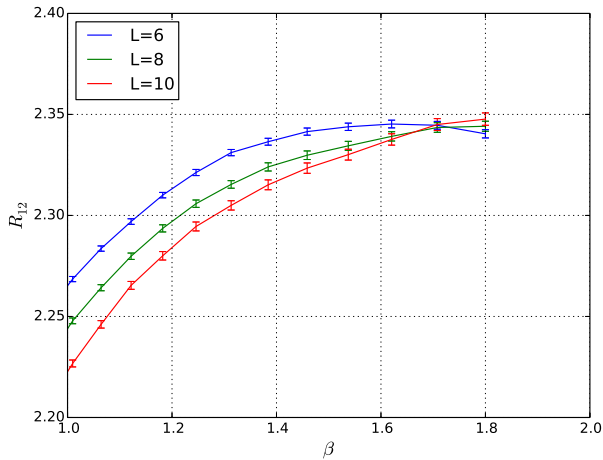}% Here is how to import EPS art
\caption{\label{fig:r12} $R_{12}$ as a function of inverse temperature for different system sizes. 
An intersection can be seen at around $T \approx 0.6$. We use the jackknife method to estimate 
the error bar from sample-to-sample variation.  
}
\end{figure}

\begin{figure}[ht]
  \includegraphics[width=0.5\textwidth]{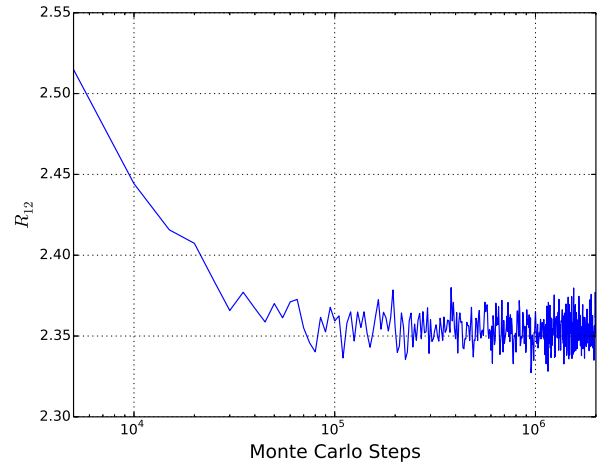}% Here is how to import EPS art
  \caption{\label{fig:MCsteps} $R_{12}$ for $L=10$ at $\beta=1.8$, as a function of the 
number of Monte Carlo sweeps.  We believe the averaged data is equilibrated 
for $10^6$  sweeps, and it passes the logarithm binning 
test \cite{Alvarez-etal-2010}. The main contribution to the error is from the realization 
averaging.}
\end{figure}

\begin{figure}[ht]
  \includegraphics[width=0.5\textwidth]{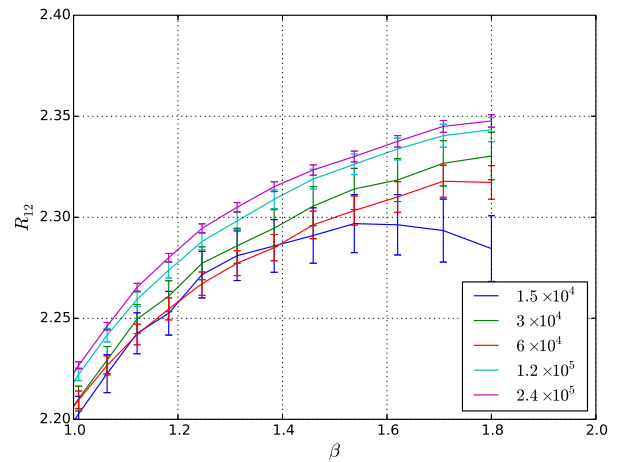}% Here is how to import EPS art
  \caption{$R_{12}$ for $L=10$ and low temperatures ($\beta \geq 1.0$). We show five different number of 
realizations from fifteen thousands to two-hundreds forty thousands.}
\label{fig:R12_l10_samples}
\end{figure}

Fig.~\ref{fig:r12} displays $R_{12}$. In contrast to other quantities, $R_{12}$ shows an intersection 
at about $T \approx 0.6$. We do not think we have sufficient data to perform a reasonably accurate finite size 
scaling analysis to report the exponent or even to quantify the correction. \cite{Hasenbusch-etal-2008}
Moreover, the data for $L=6$ does not seem to fit into a finite size scaling form with the curve bending downward.
Unfortunately, parallel tempering Monte Carlo is not robust enough for simulating larger lattices in a reasonable 
amount of time, this can be related to the temperature chaos \cite{Ritort-1994,Fernandez-etal-2013,Katzgraber-etal-2007}. The number of replicas needed to equilibrate the system also 
increases substantially as the system size increases, we already used $56$ temperature 
replicas for $L=10$ simulations. We plot $R_{12}$ versus the number of Monte 
Carlo sweeps in Fig.~\ref{fig:MCsteps}. We believe the data is sufficiently equilibrated for averaged quantities.
% as we show in this paper.
The major contribution to the error is from the limited number of disorder realizations. Fig.~\ref{fig:R12_l10_samples} shows $R_{12}$ 
for $L=10$ and different number of realizations. We clearly see that the data converges only when the number of realizations is fairly large.
This is one of the prominent hurdles of using higher momentum susceptibility as a diagnosis. 
We note that the effective one dimensional model also shows crossing behavior,
albeit the crossing points do not show a systematic trend\cite{Larson-etal-2013}.

{\it Conclusion.} In summary, we perform Monte Carlo simulations of the three dimensional
Edwards-Anderson model in a finite external field. The goal is to reexamine the long 
standing problem whether mean field behavior, specifically a spin glass phase, can
exist in such a model without time-reversal symmetry. We focus on the equilibrium quantities
of this notoriously difficult system. By taking advantage of the new commodity multi-threaded 
graphic computing units architecture we drastically reduce the computation time. 
The results for the Binder ratio and correlation length show no signal of intersection, thus they point to the absence of spin glass transition
according to conventional wisdom. On the other hand, the ratio of susceptibilities 
$R_{12}$ does show crossing behavior for relatively small system sizes ($L=6,8,10$). We did perform simulations for larger system sizes, 
but we are not confident that those simulations reach equilibrium since the data is too noisy in particular for $R_{12}$.
With the present system sizes and the statistical error bar, a rigorous data analysis does not seem to deliver unbiased information. 
This situation is rather discouraging, as simulations at this low temperature for much large system sizes 
using the present method are daunting.
This calls for a more thorough study on the model with different approaches.
%or at different perspectives. 
Possible directions include: 1) using models with continuous random distribution which are 
easier to thermalize than that with bimodal distribution; 2) analyzing the
data for the distribution of the overlap parameter, instead of average quantities. We notice a preprint before
we finished the present paper where the conditioning variate method
is used to expose the silent features from the data \cite{Baity-Jesi-etal-2014}. 

This work is sponsored by the NSF EPSCoR Cooperative Agreement No. EPS-1003897 with additional support 
from the Louisiana Board of Regents. 
Portions of this research were conducted with high performance computational resources provided by 
Louisiana State University (http://www.hpc.lsu.edu).  We thank Helmut Katzgraber and Karen 
Tomko for useful conversations.

%\nocite{*}
\bibliography{apssamp}% Produces the bibliography via BibTeX.

\end{document}